\documentclass[preprints,article,accept,moreauthors,pdftex]{Definitions/mdpi}
\firstpage{1} 
\pubvolume{1}
\issuenum{1}
\articlenumber{0}
\pubyear{2021}
\copyrightyear{2020}
\datereceived{} 
\dateaccepted{} 
\datepublished{} 
\hreflink{https://doi.org/} 

\Title{On the emergence of orientational order in folded  proteins with implications for allostery}

\TitleCitation{On the emergence of orientational order in folded  proteins with implications for allostery}

\Author{Debayan Chakraborty $^{1}$\orcidA{}, Mauro Lorenzo Mugnai $^{1}$\orcidC and D. Thirumalai $^{1}$\orcidB{}}

\AuthorNames{Debayan Chakraborty, Mauro Lorenzo Mugnai and D. Thirumalai}

\AuthorCitation{Chakraborty, D.; Mugnai, M.; Thirumalai, D.}

\address{%
$^{1}$ \quad Department of Chemistry, The University of Texas at Austin, USA}

\corres{Correspondence: dave.thirumalai@gmail.com}

\abstract{The beautiful structures of single and multi-domain proteins are clearly ordered in some fashion but cannot be readily classified using group theory methods that are successfully used to describe periodic crystals. For this reason, protein structures are considered to be aperiodic, and may have evolved this way for functional purposes, especially in instances that require a combination of softness and rigidity within the same molecule. By analyzing the solved protein structures, we show that orientational symmetry is broken in the aperiodic arrangement of the secondary structural elements (SSEs), which we deduce by calculating the nematic order parameter, $P_2$. We find that the folded structures are nematic droplets with a broad distribution of $P_2$.  We argue that non-zero values of $P_2$, leads to an arrangement of the SSEs that can resist mechanical forces,  which is a requirement for allosteric proteins.  Such proteins, which resist mechanical forces in some regions while being flexible in others, transmit signals from one region of the protein to another (action at a distance) in response to binding of ligands (oxygen, ATP or other small molecules). }


\begin{document}


\section{Introduction}

In his prescient and magnificent monograph, {\it What is Life?}, Schr{\" o}dinger~\cite{schrodinger1943life} states that, and we quote "....the most essential part of a living cell  - the chromosome fiber - may suitably be called {\it an aperiodic crystal}". He goes on to add that the difference between the well-known periodic crystal and and an aperiodic crystal is "....the same kind as that between an ordinary wallpaper in which the same pattern is repeated again and again in regular periodicity and a masterpiece of embroidery, say Raphael tapestry, which shows no dull repetition, but an elaborate, coherent, meaningful design traced by the great master", and concludes that aperiodic crystal is the "material carrier of life". Of course, similar comments are applicable to proteins, the workhorses of the cellular machinery that carry out  diverse range of functional roles, by the complex organization of their  three-dimensional structures. Because proteins can be crystallized, they diffract X-rays but their arrangement cannot be classified using a unit cell, which is used for describing the pattern repeated in the entire crystals. More importantly, a single folded globular protein (or a complex) does have some kind of order (certainly discernible to the eye) but is not periodic, and hence qualifies as an aperiodic crystal. The presence of well-defined scaffolds within the protein architecture, does not imply structural rigidity, and it is reasonable to suggest that nearly all proteins can access evolutionarily tuned deformation modes~\cite{Falke2002,Frauenfelder01PNAS,Frauenfelder91Science} in order to carry out their functions. In addition, it has been shown that coexistence of symmetrically  arranged secondary structural elements (SSEs) with flexible regions within a single protein molecule could also enhance the multiplicity of folding pathways~\cite{Klimov05JMB}. Thus, the aperiodic folds of proteins could have multiple functional consequences.  Understanding the general rules underlying the packing of SSEs~\cite{Levitt76Nature,Levitt_packing,Chothia_review_1984} is also important from the perspective of structure prediction. In a pioneering study~\cite{Levitt76Nature}, Levitt and Chothia reached far reaching conclusions by analyzing the organization of the SSEs in 31 protein structures. They suggested that $\beta$-strands that are adjacent in the sequence are ordered by forming hydrogen bonds or van der Waals contacts. It was also realized that such SSEs could form diffusing folding units, which then consolidate the native structure~\cite{Levitt76Nature}.    These ideas have been exploited in recent times in \textit{de novo} protein design~\cite{Minami2014,wood2015,Baker2014},  rationalization that geometry and symmetry considerations alone restrict the formation of native folds~\cite{Hoang04PNAS}, and in exploring the so-called `dark matter' of the protein-universe~\cite{LevittPNAS2009,DonoghuePNAS2015}.  
In a most insightful article, Wolynes~\cite{Wolynes96PNAS} recognized explicitly the importance of aperiodic crystalline nature of proteins, and the resulting approximate symmetry. Linking these ideas to the principle that the folded state should be minimally frustrated~\cite{Onuchic04COSB}, he suggested a connection between folding dynamics, approximate symmetry, and the underlying energy landscape. Thus, not only structures, but functional protein dynamics is related to the aperiodic nature of the folded state of proteins.



An intriguing function of both single folded proteins or oligomers is allosteric signaling (action at a distance), which involves signal propagation through an ordered protein, usually in response to binding of a ligand. An excellent monograph on allostery from a unique perspective has been recently been published by Rob Phillips~\cite{Phillips}. This requires the ability to impact chemistry occurring at sites beyond the reach of electrostatic forces between atoms -- in other words it is an  emergent property~\cite{Anderson:1997} of the protein structure itself. The best studied example is the structural transitions that occur in Hemoglobin~\cite{Monod65JMB,Eaton1999NSB}, when oxygen molecule binds to the heme group. Elsewhere~\cite{DT_review}, we have argued that allosteric proteins (monomers,  multi-domain proteins, or molecular machines) must be aperiodic~\cite{schrodinger1943life}. This means that these biological molecules must have at least an approximate symmetry, implying that at least a portion of the complex must be rigid~\cite{Anderson:1997}  in order to resist mechanical deformations, but not overly so, thus enabling the execution of those chemical and mechanical functions, which rely upon conformational flexibility.
Without an approximate symmetry, allosteric transitions cannot occur. Thus, there has to be a symmetry breaking transition that must occur when proteins fold from unfolded high symmetry state. Here, we suggest that there is a breakage of orientational symmetry when the proteins fold. This is manifested in the emergence of nematic order, in some instances only weakly so, which is sufficient for them to behave as allosteric systems.  Because allostery involves accessing alternate aperiodic states, which are excitations around the ground state, both the ground and excited states must have similar, but not necessarily identical symmetry. For biological viability, the free energy gap between the ground and the excited state cannot be too large~\cite{DT_review}, which also implies that the structured regions cannot be very rigid. The need for excited states implies that allosteric signaling is transmitted by "allosterons"~\cite{Mcleish2018PTRB}, much like phonons in periodic crystals.   We show, by analyzing the structures of a number of globular proteins and oligomeric assemblies, that they have non-zero values of the nematic order parameter, $P_2$. There are large variations in $P_2$. The values of $P_2$ for allosteric proteins are non-zero but not close to unity. As the magnitude of the order parameter measures the degree of the broken symmetry~\cite{Anderson:1997}, intermediate values of the order parameter reflect the delicate balance between rigidity and conformational flexibility.

\section{Methods}

\subsection{Protein Structures} 
\textbf{Database of globular proteins:} To build a representative database of globular proteins, we considered only those sequences from the PDB database that meet the following criteria: (1) mutual sequence identity of no more than 25\%. (2) at least 40 amino acids in length (3) structural resolution of at least 3,\,\AA. The list of such sequences was determined using the PISCES web-server~\cite{Wang2003}.  
Any sequence that is known to form a membrane protein has a corresponding entry in either the PDBTM~\cite{PDBTM},\textit{mpstruc}~\cite{mpstruc}, or the OPM ~\cite{OPM} database, and is excluded from our analysis.  We do so out of an abundance of caution because unlike globular proteins, membrane proteins are not evolved to be optimally compact, and the membrane itself plays an active role in their folding and assembly. 

\textbf{Subset of allosteric domains:} A non-redundant subset of experimentally determined allosteric domains was obtained from the "Core Set" of the  ASBench dataset maintained by Zhang and coworkers~\cite{ASBENCH,Huang2014}. The ASBench suite consists of 235 unique allosteric sites, predominantly found in \textit{E. Coli.} and humans. These allosteric sites are distributed among diverse protein families, including transferases, hydrolases, oxidoreductases, transcription factors, and lyases~\cite{ASBENCH}. 

\subsection{Data Analysis} 

\textbf{Residue-specific secondary structure:} The residue-wise secondary structure of the globular proteins in the curated database was determined using the DSSP algorithm~\cite{DSSP} implemented within the \textit{mdtraj} program~\cite{McGibbon2015MDTraj}.  For simplicity, we assigned the secondary structure based on a three-class system,  which identifies each residue as being in helix (H), $\beta$-sheet (E), or in coil/turn configuration (C).  Based on the secondary structure content, we segregate all the proteins in the database into three types: if the overall $\alpha$-helicity of a protein structure exceeds 20\%, and the $\beta$-content is less than 10\%, the protein is classified as $\alpha$-rich; if the net $\beta$-content is greater than 20\%, and the $\alpha$-helix content is less than 10\%, the protein is considered to be $\beta$-rich;  If both the $\alpha$-helix and the $\beta$-content exceeds 12\%, then we classify the protein as $\alpha+\beta$. We note that the cutoffs employed for determining the different protein structural classes are quite robust, and similar statistics were obtained for variations within a reasonable range of these cutoff values. 
To further distinguish among the different topologies adopted by the three protein classes, we use the CATH classification of proteins~\cite{CATH} as a guide.

 \textbf{Ordering of secondary structure elements (SSE):} To probe the nematic ordering of secondary structure elements (SSE) within the protein fold, we compute the nematic order parameter using the WORDOM package~\cite{WORDOM,wordom2}.  The traceless second-rank tensor $Q$, can be computed from molecular vectors, $\vec{w_{i}}$ as~\cite{Wilson1996,nematic_Andrienko}: 
 
 \begin{equation}
 Q_{mn} = \frac{1}{N_{s}} \sum_{i=1}^{N_{s}} \frac{3}{2} w_{im}w_{in} - \frac{1}{2} \delta_{mn}.
 \end{equation}
 \noindent In Eq.~(1), indices $m$ and $n$ denote the $x$, $y$ and $z$ directions, and $\vec{w}_{i}$ denote the molecular vectors connecting the $C_{\alpha}$ atoms located at the terminii of secondary structure elements, such as the $\alpha$-helix, or the $\beta$-sheet; $N_{s}$ denotes the total number of $\alpha$-helices and $\beta$-sheets within the protein fold; these secondary structure elements (SSE) can be part of a single or multiple different chains.  Diagonalization of $Q$ yields the nematic director, $d$, which is the eigenvector associated with the maximum nonzero eigenvalue~\cite{Frenkel84,Wilson1996}.


From $Q$ (Eq.~1), and the director $\vec{d}$,  the nematic order parameter, $P_{2}$ is computed as~\cite{wordom2}:  
\begin{equation}
P_{2}  =  \vec{d} \cdot Q \cdot \vec{d}
\end{equation}
\label{nematic}
For a near perfect ordering of SSE, $P_{2} \approx 1$, while $P_{2} \approx 0$ for isotropic arrangements. 

A diagrammatic illustration of the vectors, $\vec{w}_{i}$, in the context of a $\alpha+\beta$ protein is shown in Figure 1. Here, $\vec{h}$ denotes the $\vec{w_{i}}$ vectors directed along the axis of the $\alpha$-helical segment, and $\vec{b}$ denotes the $\vec{w_{i}}$ vectors directed along the an extended $\beta$-sheet in the protein.

\begin{figure}
\centering
\includegraphics[width=0.30\textwidth]{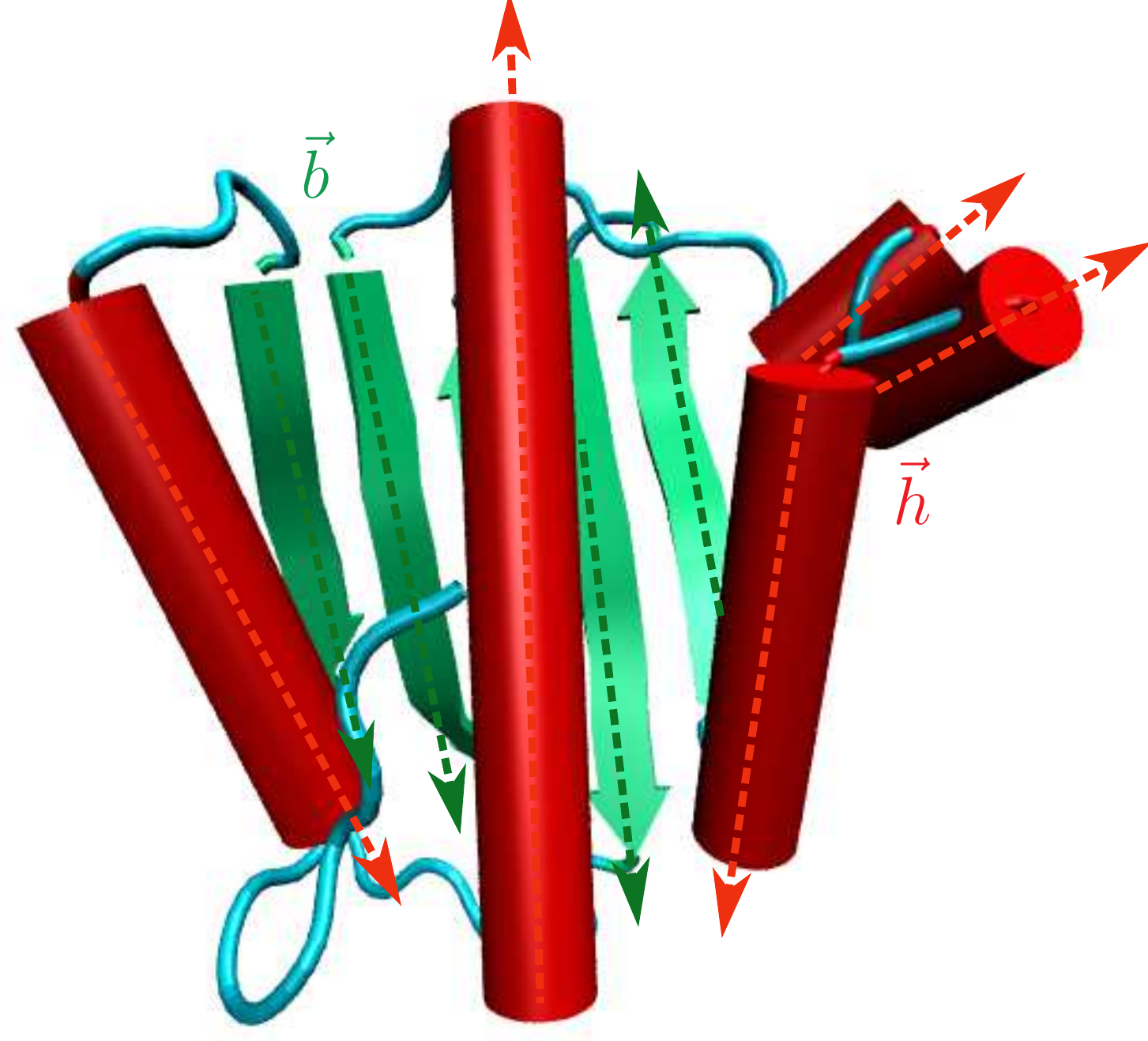}
\caption{An illustration of the vectors, $\vec{w_{i}}$ that are used to compute the nematic order parameter, $P_{2}$ (see Eq.~2), for a $\alpha+\beta$ protein. The protein is shown in a cartoon representation, with the $\alpha$-helices depicted as red cylinders, and the extended $\beta$-sheets colored green. The irregular coils/turns connecting the different secondary structure elements (SSE) are in cyan. The vectors (for example, $\vec{h}$) connecting the $C_{\alpha}$ atoms located at the termini of   $\alpha$-helices are depicted as red arrows.  Green arrows  denote the vectors (for example, $\vec{b}$) connecting the $C_{\alpha}$ atoms located at the termini of extended $\beta$-sheets.  }
\end{figure}


\section{Results}

\textbf{Single- and multi-chain proteins have different organization of secondary structural elements (SSE):} The distribution of the nematic order parameter, $P_{2}$, for the subset of single and multi-chain globular proteins that meet the structural and sequence-identity criteria (see Methods), are shown in  Figure 2(a). As expected, the $P_{2}$ distribution for the multi-chain proteins peaks at relatively lower values  compared to the single-chain proteins, due to the additional orientational degree of freedom dictating the relative arrangement in space of different monomeric units.   In the aperiodic states, $P_2 \ne 0$, implying the ordered states are indeed nematic droplets in which orientational symmetry is broken. The relatively higher probability density in the region, $0.4 \leq P_{2} \leq 0.7$, for single-chain proteins, further indicates that they adopt more folds/topologies commensurate with intermediate ordering of the SSEs. The multi-chain $P_{2}$ distribution, on the other hand,  features slightly enhanced probabilities in the tail region, beyond $P_{2}\approx 0.8$, suggesting that highly ordered assemblies (with only few packing defects) are more likely to result from oligomerization, rather than self-association of multiple SSE within single-chains.


\begin{figure}[!htbp]
\hspace*{-2cm}
\centering
\includegraphics[width=0.80\textwidth]{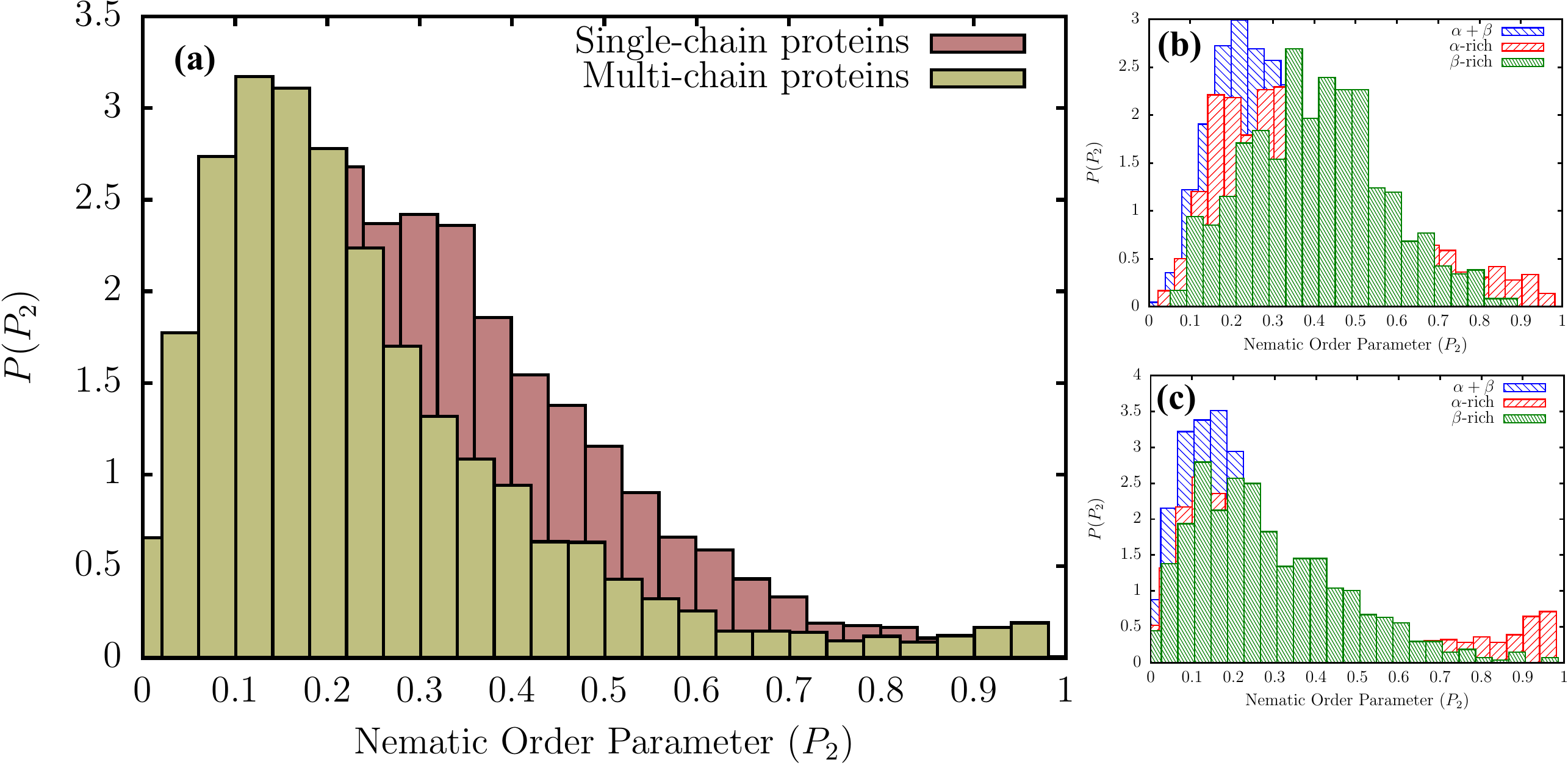}
\caption{(a) Distribution of the nematic order parameter, $P_{2}$, for single- and multi-chain globular proteins present in the database constructed using the PISCES web-server. The secondary-structure dependent $P_{2}$ distributions in single- and multi-chain proteins are shown in (b) and (c), respectively. As outlined in the Methodology section, the residue-dependent secondary-structures were computed using the DSSP algorithm.}
\end{figure}

Additional insights into the different folds/topologies that are adopted by single- and multi-chain proteins can be obtained from the secondary-structure dependent $P_{2}$ distributions, shown in Figure~2(b) and Figure~2(c). We classify all proteins in the database into three broad classes: $\alpha$-rich, $\beta$-rich and $\alpha + \beta$ (see Methods section).  Among $\alpha$-rich proteins, the ordering of SSE is maximized within helical bundles and coiled-coils, featuring the helix-loop-helix type topology.  A representative example of this fold includes the seven-helix bundle periplasmic sensor domain of the TorS receptors found in \textit{E. Coli.} (Figure 3). 
For single-chain $\beta$-rich proteins, the probability density at $P_{2} \geq 0.8$ is lower than the $\alpha$-rich counterparts, suggesting that the alignment of SSEs relative to one another is somewhat less efficient. The ordering of SSEs within single-chain $\beta$-rich proteins is maximized primarily in sandwich domains having the immunoglobin fold (involved in passive muscle elasticity), although some highly ordered structures also exhibit other topologies, such as jelly rolls,  and $\beta$-prisms. A few representative examples of such folds are shown in Figure~3. 

 In contrast to the other two protein classes, single-chain $\beta$-proteins seem to form a large number of folds characterized by intermediate ordering of the SSEs ($P_{2}$ in between 0.4 and 0.5). We find that these folds largely have the sandwich type architecture, with either the immunoglobin or the jelly roll type topology. Some proteins also adopt the more exotic $\beta$-propeller topology, with the number of blades varying between five and eight (Figure~3).  The difference in the $P_{2}$ distributions between single-chain and multi-chain proteins can largely be explained by this ability of single-chain $\beta$-rich proteins to form such diverse folds, exhibiting intermediate ordering of SSEs.   As is evident from Figure~2(b), $\alpha + \beta$ proteins are also less likely to form folds that maximize the ordering of the SSEs ($P_{2} \geq 0.8$) as compared to $\alpha$-rich proteins.  Nonetheless, we find few topologies, such as $\alpha\beta$-plaits, leucine-rich repeats, where SSEs could be maximally aligned (Figure 3). The peak of the $P_{2}$ distribution for $\alpha + \beta$ proteins occurs at $\approx$ 0.2, consistent with a large number of architectures found in the database, such as the two-layer sandwich, three-layer aba sandwich, $\alpha\beta$-complex, and the $\alpha\beta$-barrel. In particular, we find that the Rossman fold (three-layer aba sandwich), $\alpha\beta$-plaits, and TIM barrels ($\alpha\beta$-barrel) are particularly favored (Figure 3). 

\begin{figure}[!htbp]
\hspace*{-2cm}
\centering
\includegraphics[width=0.60\textwidth]{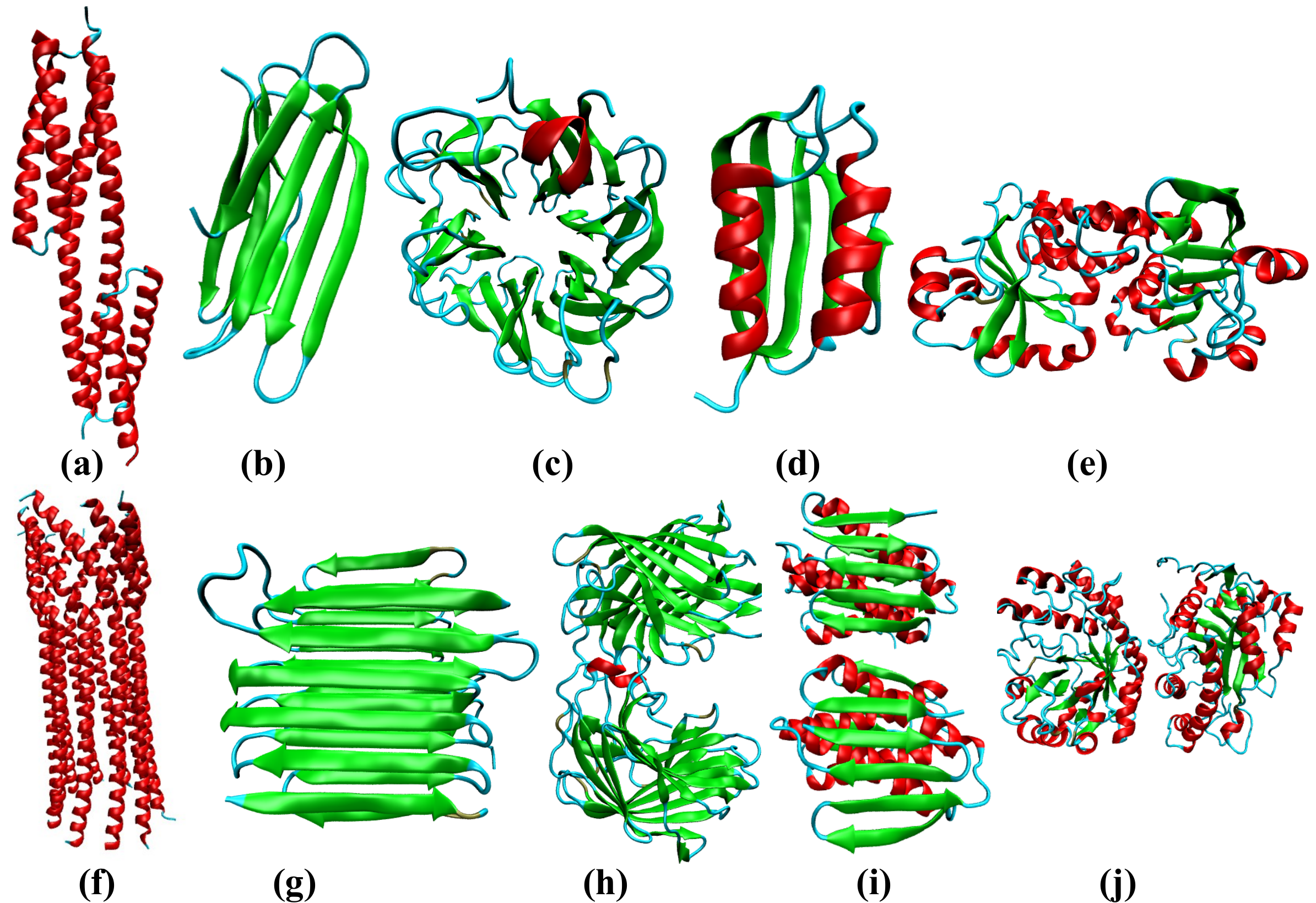}
\caption{Some representative protein folds found among proteins in the PDB database. (a) seven-helix bundle periplasmic sensor domain of the TorS receptors (PDB ID: 3I9Y). (b) Immunoglobin-like fold (PDB ID: 2XSK) present in the \textit{E. Coli.} curli protein  (c) 6-blade $\beta$-propeller (PDB ID: 3A72) found within a high-resolution crystal structure of Penicillium chrysogenum alpha-L-arabinase in complex with arabinobiose. (d) $\alpha\beta$ plait (PDB ID: 6CBU) topology found within an immunogen. (e) Rossmann fold (PDB ID: 4LJS) present in the periplasmic binding protein. (f) Central domain of bacteriophage phiX174 H protein formed by the assembly of ten $\alpha$-helices (PDB ID: 4JPN). (g) 2-solenoid topology found in antifreeze proteins (PDB ID: 4DT5) (h) jelly-roll  (PDB ID: 4HFS) found in a protein from Bacilus subtilis subsp. subtilis str. 168. (i) $\alpha\beta$-plait (PDB ID: 1S12) within the protein TM1457 from Thermotoga maritama. (j) TIM barrel (PDB ID: 3CBW) found in the YdhT protein from Bacilus subtilis. }
\end{figure}

Similar to their single-chain counterparts, multi-chain $\alpha$-rich proteins also form a larger number of orientationally ordered structures, compared to the other two protein classes, by exploiting the helix-bundle architecture, as in the case of the central domain of bacteriophage phiX174 H protein.  Multi-chain $\beta$-proteins maximize the ordering of SSEs through a diverse range of folds including aligned prisms, jelly rolls, solenoids, and immunoglobin-like sandwich architectures.  However, unlike their single-chain counterparts, the subset of multi-chain $\beta$-proteins also includes highly ordered structures ($P_{2} \geq  0.9$), such as amyloid fibrils, and antifreeze proteins exhibiting the 2-solenoid topology (Figure 3). Interestingly, multi-chain $\beta$ proteins do not adopt as many folds of intermediate order,  suggesting that additional symmetry requirements could impose constraints  on the formation  of such topologies (Figure 2(c)).  The $P_{2}$ distribution for multi-chain $\alpha + \beta$ proteins is quite similar to that for single-chains. A large number of the structures exhibit intermediate ordering of the SSEs (peak at $\approx$ 0.2), and belong to the Rossman fold, $\alpha\beta$-plait, or the TIM barrel superfamily.   In addition, like single-chains, multi-chain $\alpha + \beta$ proteins also primarily exploit the $\alpha\beta$-plait topology to maximally align SSEs (Figure 3). 

Overall, we find that for most single- and multi-chain proteins found in the PDB database, the arrangement of SSEs exhibit orientational order. It may be argued that they are of weak nematic order because $P_2$ values are not too close to unity. 
Only a few topologies are exploited to form highly symmetrical or ordered structures. 

\begin{figure}[!htbp]
\hspace*{-2cm}
\centering
\includegraphics[width=0.60\textwidth]{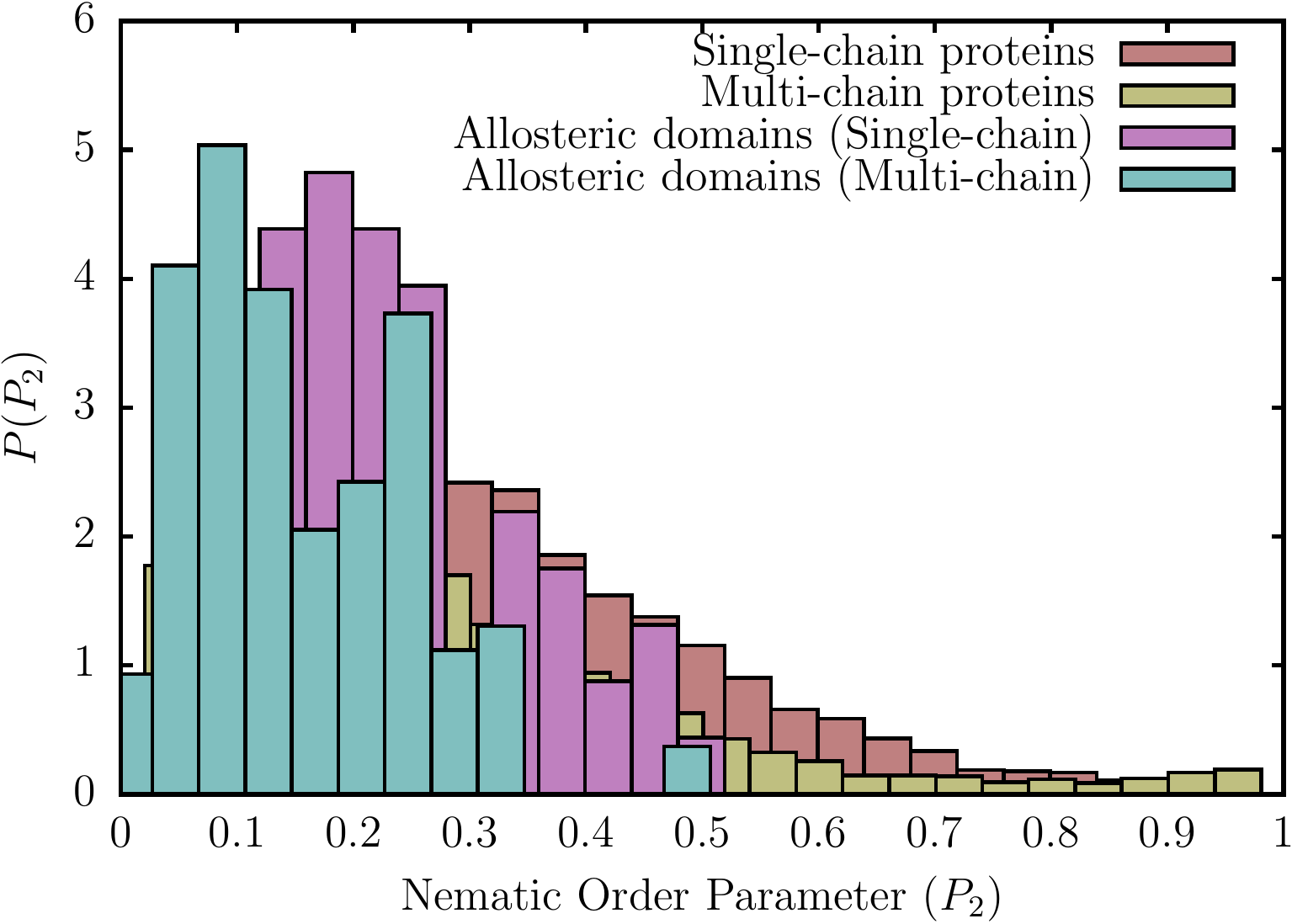}
\caption{Distribution of the nematic order parameter, $P_{2}$ for single- and multi-chain proteins culled using the PISCES web-server, and allosteric domains present in the ASBench database. For allosteric domains, the nematic order parameter,$P_{2}$, has an upper bound of $\approx$ 0.5, and the corresponding distributions lack long tails  at high $P_{2}$ values (resulting from highly ordered assemblies).}
\end{figure}

\textbf{Allosteric domains display weak nematic order:}  In most proteins, allostery is  modulated through the concerted motions of structural elements, such as strands and helices, orchestrated by flexible loops or hinges. One of the key requirements in allostery is that certain regions of the protein must be stiff enough to resist mechanical stress,  and yet the overall architecture should be flexible enough to accommodate subtle changes in conformation~\cite{DT_review}.  Therefore, understanding the precise ordering of SSEs, as well as fold preferences within allosteric domains becomes crucial.  

\begin{figure}[!htbp]
\hspace*{-2cm}
\centering
\includegraphics[width=0.60\textwidth]{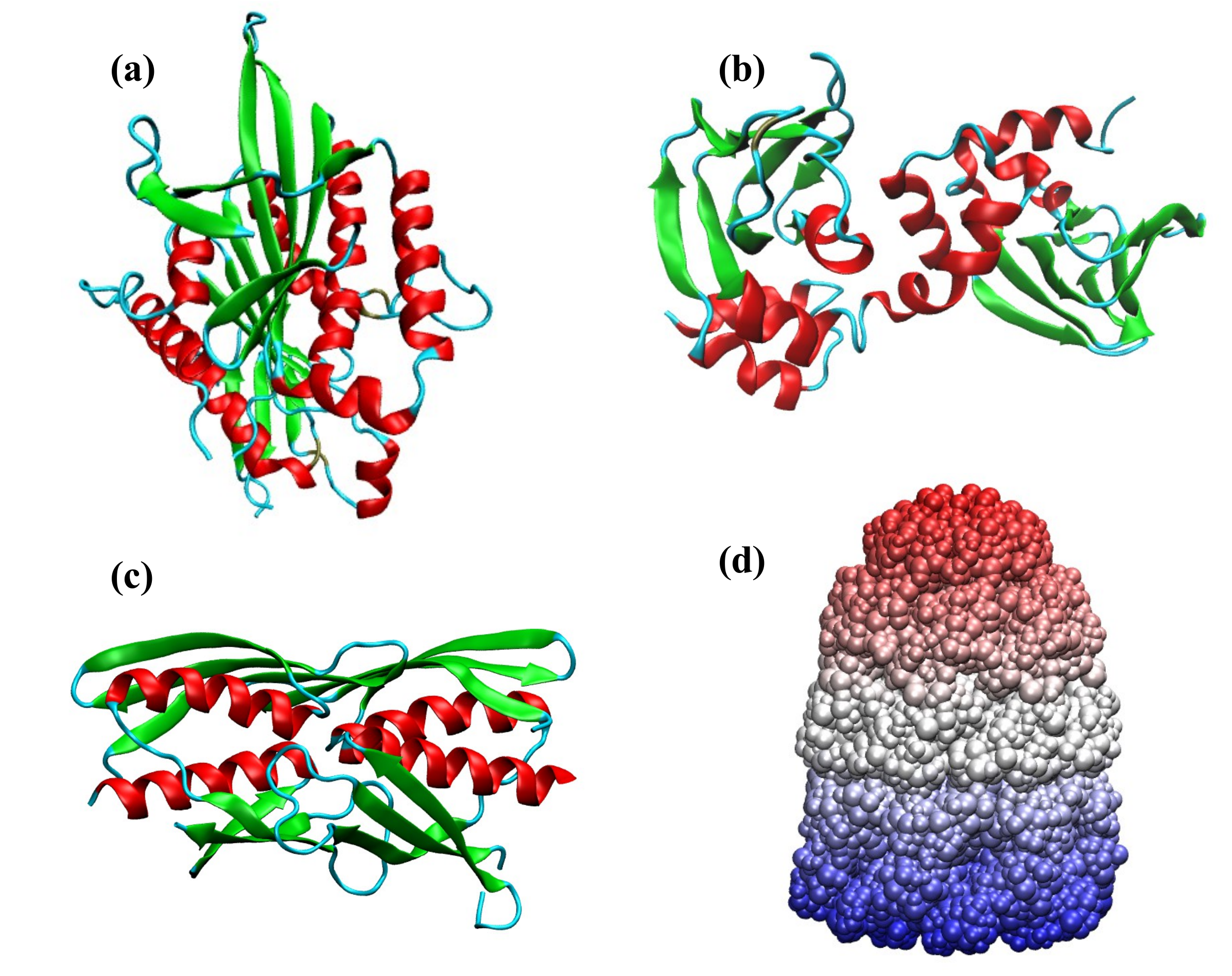}
\caption{Some representative examples of allosteric domains. (a) Human kinesin spindle protein, Eg5 (PDB ID: 3ZCW) (b) PKA regulatory subunit of yeast (PDB ID: 3OF1) (c) Regulatory domain of LiCMS (PDB ID: 3F6G) (d) Bullet shaped GroEL-GroES complex (PDB ID: 1PCQ).}
\end{figure}

We find that in contrast to protein structures culled using the PISCES web-server, the nematic order parameter, $P_{2}$ for allosteric domains does not exceed $\approx$ 0.5 (Figure~4). Therefore, the arrangement of SSEs within allosteric domains cannot be overtly stiff,  like in ordered assemblies, such as fibers and helix bundles, and must be flexible enough to access excited state  in order to accommodate structural perturbations.  In a recent study, Lai and coworkers~\cite{Xie2020} argued that allosteric domains  have a preference for certain topologies like the Rossmann fold, while certain folds like the immunoglobin-like $\beta$-sandwiches are disfavored.  Our results, support such a viewpoint, because we find that the distributions of the nematic order parameter for allosteric domains lack the long tails at high $P_{2}$ values, which in the case of single-chain $\beta$-rich proteins primarily result from immunoglobulin-like folds, which can withstand large mechanical forces. Furthermore, the enhanced  probabilities at $P{2} \approx 0.2$ (Figure~4), also reflect the preference of allosteric domains for Rossmann folds~\cite{Xie2020}. 

Among single-chain allosteric domains included within the ASBench database, the human kinesin spindle protein (also known as Eg5) exhibits the maximal ordering of secondary structural elements, with a $P_{2}$ value of $\approx$ 0.50 (Figure~5a).  Within the kinesin superfamily, Eg5 is one of the most advanced drug targets, and most designed inhibitors, such as ispinesib, specifically bind to an allosteric pocket located around 10\,\AA\, away from the ATP binding site~\cite{Eg5,Eg52}. Recent evidence~\cite{Eg52,Eg53} suggests that the binding of ispinesib and other drugs to the allosteric pocket of Eg5 only induces minimal structural changes in the arrangement of the SSEs, and much of the perturbation is accommodated by the flexible loop region.  Another prototypical single-chain allosteric domain is found in the protein kinase A (PKA) regulatory subunit of yeast~\cite{yeast1}. The enzyme consists of two cyclic adenosine monophosphate (cAMP) binding domains (Figure~5b),  and the unique interface formed between the two domains modules the allosteric mechanism underlying PKA activation~\cite{yeast1}.  Substantial structural changes, which occur at the interface due to cAMP binding,  are primarily driven by the movements of the helical subunits within the CNB domains.  Thus, in contrast to Eg5, allosteric transitions in this protein requires a very labile network of SSEs, associated with low nematic order.  We find that this is precisely the case, and the $P_{2}$ value for the yeast regulatory subunit is only around 0.15.  
   
Within the subset of multi-chain allosteric domains, the regulatory domain of Leptospira interrogans citramalate synthase (LiCMS) is associated with the highest value of $P_{2}$ ($\approx 0.5$). The functional form of LiCMS is a homodimer (Figure~5c) , and it is involved in the  catalysis of  the first step of the isoleucine biosynthesis pathway in \textit{L. interrogans}. The reaction is regulated through a feedback inhibition mechanism mediated by the binding of isoleucine to the allosteric site at the dimer interface~\cite{licms}. Structural and biochemical evidence strongly suggest that the inhibition of LiCMS is likely to occur via interdomain communication between the regulatory and catalytic domains, rather than large-scale rearrangements in the quaternary structure of the protein~\cite{licms}. This scenario for allosteric regulation, seems consistent with the somewhat rigid packing of SSEs found in this protein. As a final example, we consider the well-studied GroEL-GroES complex\cite{groEL}, an ATP-fueled  nanomachine, which aids in protein folding, by undergoing a series of complex allosteric transitions~\cite{groel_Karplus,groel_DT}.  The different rings of the GroEL-GroES complex move in a concerted fashion, with the inter-ring motions being anticorrelated,  resulting in a form of nested cooperativity~\cite{nested_coop}. It is evident that intra- and inter-subunit rearrangements underlying such complex allostery can only be attained through an arrangement of structural elements, which can be disrupted by moderate strain. Indeed, in the GroEL-GroES complex, the extent of nematic order is small ($P_{2} \approx 0.06$). 

The examples above suggest that the extent of nematic order found in allosteric proteins has been tuned to carry out a specific function. While allosteric inhibition could be accommodated by a rigid arrangement of SSEs, complex allosteric transitions require a somewhat flexible network of structural elements that are arranged in an aperiodic manner, but responds to stresses readily.  It is worth remembering that allostery requires that the two states (usually referred to as $T$ and $R$) be readily accessible upon ligand binding (oxygen to Hemoglobin (Hb) for example). In the absence of the ligand the $R$ state would be an excited state separated from the $T$ state by a free energy gap. During the $T \rightarrow R$ transition, although the overall symmetry might be preserved the extent of order often changes, reflecting the effect of the ligand. Indeed, in a family of Hemoglobins we find that $P_2$ is greater in the $R$ state compared to the $T$ state. Thus, the strict preservation of the approximate symmetry in the ground state is neither required nor is observed.   

\section{Conclusion}
 
The precise arrangement of secondary structure elements within protein structures holds important cues for understanding their potential as an aggregation seed, allosteric scaffold, or a therapeutic target, not mention the designability of proteins and their folding.  In this study, we assessed the organization of secondary structure elements for a subset of proteins found in the PDB database,  in terms of the nematic order parameter, $P_{2}$, which has been traditionally exploited in the studies of anisotropic molecules that break orientational symmetry when close packed.  We find that the distributions of $P_{2}$ are sensitive to the number of polypeptide chains present in the protein molecule, as well as the overall secondary structure.  The secondary structure elements are maximally aligned in highly ordered structures, such as helix-bundles, amyloid fibers, and antifreeze proteins.  However, as indicated by the small to intermediate values of $P_{2}$, in most allosteric proteins the arrangement of SSEs exhibits nematic order but coexists with other more flexible regions (loops for example). Such an arrangement, observed in diverse fold topologies, ensures that the aperiodically arranged structures can transmit allosteric signals over the length of the complex, while the flexible regions can rearrange to readily access the excited states for carrying out specific functions. 
 
In allosteric domains, the SSEs cannot be maximally aligned (i.e. $P_{2} \approx 1$) because that would make the structure too rigid.   The extent of nematic ordering is functionally tuned, and depends on the precise structural requirements underlying the allosteric transitions.  Based on the large overlap between the $P_{2}$ distributions near the peak regions for proteins in the PISCES database, and the ASBench subset, we speculate that many more non-fibrous proteins, than currently known,  could potentially be allosteric under suitable conditions.


\authorcontributions{D.C., M.L.M. and D.T. designed research; D.C. performed research and analyzed data; and D.C., M.L.M. and D.T. wrote the manuscript.} 

\acknowledgments{We are grateful to Bill Eaton, Gilard Haran, and Amnon Horovitz for useful discussions. This work was supported by the National Science Foundation (CHE19-00093) and by the Collie-Welch Chair administered through the Welch foundation (F-0019).} 

\conflictsofinterest{The authors declare no conflict of interest.}



\reftitle{References}

\end{paracol}
\end{document}